\def\real{{\tt I\kern-.2em{R}}}
\def\nat{{\tt I\kern-.2em{N}}}
\def\realp#1{{\tt I\kern-.2em{R}}^#1}
\def\natp#1{{\tt I\kern-.2em{N}}^#1}
\def\hyper#1{\ ^*\kern-.2em{#1}}

\def\hyperrealp#1{{\tt ^*{I\kern-.2em{R}}}^#1} 
\def\hypernat{{^*{\nat }}}
\def\hypernatp#1{{{^*{{\tt I\kern-.2em{N}}}}}^#1} 
\def\eskip{\hskip.25em\relax}

\def\Hyper#1{\hyper {\eskip #1}}
\def\leaderfill{\leaders\hbox to 1em{\hss.\hss}\hfill}
\def\srealp#1{{\rm I\kern-.2em{R}}^#1}

\def\power#1{{{\cal P}(#1)}}
\def\iff{\leftrightarrow}
\def\qed{{\vrule height6pt width3pt depth2pt}\par\medskip}
\def\pars{\par\smallskip}
\def\parm{\par\medskip}
\def\r#1{{\rm #1}}
\def\b#1{{\bf #1}}
\def\ref#1{$^{#1}$}

\def\m@th{\mathsurround=0pt}
\def\rightarrowfill{$\m@th \mathord- \mkern-6mu \cleaders\hbox{$\mkern-2mu 
\mathord- \mkern-2mu$}\hfil \mkern-6mu \mathord\rightarrow$}
\def\leftarrowfill{$\mathord\leftarrow
\mkern -6mu \m@th \mathord- \mkern-6mu \cleaders\hbox{$\mkern-2mu 
\mathord- \mkern-2mu$}\hfil $}
\def\noarrowfill{$\m@th \mathord- \mkern-6mu \cleaders\hbox{$\mkern-2mu 
\mathord- \mkern-2mu$}\hfil$}
\def\orgate{$\bigcirc \kern-.80em \lor$}
\def\andgate{$\bigcirc \kern-.80em \land$}
\def\inverter{$\bigcirc \kern-.80em \neg$}

\def\id{\par\hangindent2\parindent\textindent}
\def\textindent#1{\indent\llap{#1}}
\magnification=\magstep1
\tolerance 10000
\baselineskip  12pt
\hoffset=.25in
\hsize 6.00 true in
\vsize 8.85 true in
\font\eightrm=cmr9
\centerline{\bf An Ultralogic Unification for All Physical Theories}\par\bigskip 
\centerline{Robert A. Herrmann}\parm
\centerline{Mathematics Department}
\centerline{U. S. Naval Academy}
\centerline{572C Holloway Rd.}
\centerline{Annapolis,  MD 21402-5002}
\centerline{8 JAN 2001}\bigskip
{\leftskip=0.5in \rightskip=0.5in \noindent {\eightrm {\it Abstract:} In this paper, the set of all physical theories is represented by a countable collection of consequence operators $\rm \{S^V_{N_j}\mid j \in \nat \}$ defined on a language ${\Lambda}.$ It is established that in the Grundlegend Structure, a nonstandard structure, there exists an injection $\cal S$ such that for any significant natural-system representation $\rm W \subset {\Lambda},$ ${\cal S}_{\rm W}$ is an ultralogic such that $\bigcup\{\b S^{\b V}_{\b N_j}(\b W)\mid j \in \nat \} = {\cal S}_{\rm W}(\Hyper {\b W})\cap {\bf \Lambda}.$   
 \par}}\par\bigskip
\noindent{\bf 1. Introduction.}\parm
Seventy years ago, Tarski (1956, pp. 60-109) introduced the mathematical object called a {\it consequence} operator as a model for various aspects of deductive thought. There are two such mathematical theories investigated, the {\it general} and the {\it finite} consequence operators (Herrmann, 1987). The finite consequence operators are usually the operators that model human thought processes that use but finite arguments and a finite collection of premises to arrive at a specific conclusion. Let $\r L$ be a nonempty language, $\cal P$ be the power set operator and $\cal F$ the finite power set operator. \parm
{\bf Definition 1.1.} A mapping $\r C\colon \power{\r L} \to \power {\r L}$ is a general consequence operator (or closure operator) if for each $\r X,\ \r Y \in 
\power {\r L}$\pars
\indent\indent (i) $\rm X \subset C(X) = C(C(X)) \subset L$; and if\pars
\indent\indent (ii) $\rm X \subset Y$, then $\rm C(X) \subset C(Y).$\pars
\noindent A consequence operator C defined on L is said to be {\it finite} ({\it finitary}, or {\it algebraic}) if it satisfies\pars
\indent\indent (iii) $\rm C(X) = \bigcup\{C(A)\mid A \in {\cal F}(\r X)\}.$\par\medskip
{\bf Remark 1.2.} The above axioms (i), (ii), (iii) are not independent. Indeed, 
(i), (iii) imply (ii). Hence, the finite consequence operators defined on a specific language form a subset of the general operators.\parm
Natural-systems are named and defined by scientific disciplines. Each is an arrangement of named physical objects that are so related or connected as to form an identifiable unity. Except for the most basic, natural-systems always require the existence of accepted natural laws or processes for, at least, two events to occur. It is required that a natural-system either be constructed by application of natural laws or processes from more fundamental physical objects (natural-systems); or that the natural-system is altered in its development by such natural laws or processes, in which case the original natural-system may be considered as a more fundamental physical object. \pars 
Explicit statements for a natural law or process and the theories they yield are human inventions that imitate, in the words of Ferris (1979, p. 152), intrinsic natural laws or processes that govern the workings of those portions of our universe that are comprehensible. Individuals apply various mental processes to a set of hypotheses that include a set of natural laws or processes and predict behavior for a natural-system. Mental processes are also applied to natural laws or processes in order to construct our material ``man made universe.'' Consequence operators model such mental behavior. Indeed, these operators model many general mental processes not merely the standard notion termed as ``deduction.'' \parm
\noindent {\bf 2. Axiomatic consequence operators.}\parm
Prior to simplification, we need to assume that our consequence operators are axiomatic, where the axioms include appropriate natural laws or processes. Also, we need the fundamental philosophy of modern science that,  with the exception of the accepted and most fundamental of physical objects,  all named natural-systems are obtained by application of natural laws or processes to physical objects that are defined as more fundamental in character than the natural-systems of which they are constituents. Obviously, specified natural laws or processes alter specific natural-system behavior. As mentioned, the results in this paper are not restricted to what is usually termed as deduction. As done in Herrmann (1999, p. 12), we only consider equivalent representatives as the members of $\r L$. (This is not the same notion as consequence operator logical equivalence.) Let $\rm {\cal C}(L)$ [resp. $\rm {\cal C}_f(L)$] be the set of all general [resp. finite] consequence operators defined on $\r L,$ where $\r A \subset \r L$ is the set of logical axioms for $\r F \in \rm {\cal C}(L)$ [resp. $\rm {\cal C}_f(L)$].\pars 

Although, usually,  such consequence operators are considered as axiomatic,  in this application the use of axiomless operators (Herrmann  1987, p. 3) leads to a significant simplification. For $\r F \in \rm {\cal C}(L)$ [resp. $\rm {\cal C}_f(L)$], let $\r A \cup \r N\subset \r L$ and suppose that  $\r F(\emptyset)\supset \r A \cup \r N.$ (Note: $\r N$ does not denote the natural numbers). Then,  $\emptyset \subset \r A \cup \r N$ yields $\r F(\emptyset) \subset \r F(\r A \cup \r N),$ and $\r A \cup \r N \subset \r F(\emptyset)$ yields that   $\r F(\r A \cup \r N) \subset \r F(\r F(\emptyset)) = \r F(\emptyset)$. Hence,  $\r F(\emptyset) = \r F(\r A \cup \r N).$ Further,  note that if $\r B \subset \r A \cup \r N,$ then since $\emptyset \subset \r B,$ it follows that $\r F(\emptyset) = \r F(\r A \cup \r N)\subset \r F(\r B)\subset \r F(\r F(\r A \cup \r N)) = \r F( \r A\cup \r N)$ and $\r F(\r B) = \r F(\r A\cup \r N).$  The objects in $\r F(\r A \cup \r N)$ behave as if they are axioms for $\r F.$ Can we use this axiomatic behavior to generate formally a specific consequence operator $\r C,$ where $\r C(\emptyset) = \emptyset$ and the only results displayed by this model are conclusions not members of $\r F(\r A \cup \r N)$? If such a meaningful consequence operator exists,  then this approach is acceptable since if natural laws or processes, as represented by $\r N,$ are stated correctly,  such as always including any physical circumstances that might restrict their application,  then they behave like physical ``tautologies'' for our universe. For such a basic consequence operator $\r F,$ the set $\r F(\emptyset)$ is composed of all of the restatements of $\r N$ that are considered as ``logically'' equivalent,  and all of the pure ``logical'' theorems.\pars 

 In general,  various forms of scientific argument are modeled by consequence operators,  where the use of axioms is a general process not dependent upon the axioms used. The axioms are but inserted into an argument after which the actual rules of inference are applied that might yield some $\r x \in {\r L} - \r F(\emptyset)$. It is this $\r x$ that may yield something not trivial. In the physical case,  this $\r x$ may represent some aspect of an actual physical object distinct from the natural laws or processes.\parm
\noindent {\bf 3. Rules that generate consequence operators.}\parm
In this investigation,  the term ``deduction'' is broadly defined. Informally,  the pre-axioms $\r A\cup \r N$ is a subset of our language ${\r L},$ where $\r N$ represent natural laws or processes, and there exists a fixed finite set ${\bf RI} =\{\rm  R_1,\ldots, R_p\}$ of n-ary relations $(\rm n \geq 1)$ on ${\r L}.$ The term ``fixed'' means that no member of $\bf RI$ is altered by any set $\rm X$ of hypotheses that are used as discussed below. It is possible, however, that some of these $\rm R_i$ are $\r N$ dependent. It can be effectively decided when an $\r x \in \r L$ is a member of $\r A\cup \r N$ or a member of any of the fixed 1-ary relations.  Further,  for any finite $\r B \subset \r L$ and an $(j +1)$-ary $\r R_{\r i} \in {\bf RI},\ \r j > 1$ and any $\r f \in \r R_{\r i},$ it is always assumed that it can be effectively decided whether the k-th coordinate value $\r f(\r k) \in \r B,\ \r k= 1,\ldots,\r j.$ It is always assumed that a mental or equivalent activity called {\it deduction} from a set of hypotheses can be represented by a finite (partial) sequence of numbered (in order) steps $\r b_1,\ldots,\r b_{\r m}$  with the final step $\r b_{\r m}$ the conclusion of the deduction. All of these steps are considered as represented by objects from the language $\r L.$ Any such representation is composed either of the zero step,  indicating that there are no steps in the representation,  or one or more steps with the last numbered step being some $\r m >0$. In this inductive step-by-step construction,  a basic rule used to construct this representation is the {\it insertion} rule. If the construction is at the step number $\r m \geq 0,$ then the insertion rule,  {\bf I},  is the ``insertion of an hypothesis from $\r X \subset {\r L},$ or insertion of a member from the set $\r A\cup \r N,$ or the insertion of any member of any 1-ary relation,  and denoting this insertion by the next step number.'' If the construction is at the step number $\r m > 0,$ then the {\it rules of inference},  {\bf RI},  are used to allow for an insertion of a member from $\r L$ as a step number $\r m+1,$ in the following manner. For any 
$(\r j+1)$-ary $\r R_{\r i} \in {\bf RI},$ $\rm 1\leq j,$ and any $\r f \in \r R_{\r i},$ if $\r f(\r k) \in \{\r b_1,\ldots, \r b_{\r m}\},\ \r k=1,\ldots,\r j,$ then $\r f(\r j+1)$ can be inserted as a step number $\r m+1.$ Note, in particular, how specific ``choices'' are an essential part of the process here termed as deduction. The deduction is constructed only from the rule of insertion or the rules of inference as here described. \pars 
It is not difficult to show that if you apply these procedures to obtain the final step as your deduction,  then these procedures are modeled by a finite consequence operator. For the language $\r L,$ a set of pre-axioms $\r A\cup \r N,$ a set {\bf RI} and any $\r X \subset {\r L},$ define the set map $\rm C_N,$ by letting $\rm C_N(X)$ be the set of {\bf all} members of $\r L$ that can be obtained from $\r X$ by ``deduction.'' Clearly,  by insertion
$\rm X \subset  C_N(X).$ Since $\rm C_N(X) \subset {L},$ then we need to consider the result $\rm C_N(C_N(X))$. Since no member of the set $\bf RI$ is altered by introducing a different set of hypotheses such as $\rm C_N(X),$ then this composition is defined. Let $\rm x \in C_N(C_N(X)).$  By definition,  $\r x$ is the final step in a finite list $\{\r b_{\r i}\}$ of members from $\r L.$ The steps in this finite ``deduction'' from which $\rm x \in L$ is obtained are the $\b I$ steps, where we only added to this insertion members of $\rm C_N(X),$ and the {\bf RI} steps, as defined above, where the {\bf RI} are fixed. Suppose that $\rm b_{i}\in C_N(X)$ is any of these additional insertions. Simply construct a new finite sequence of steps by substituting for each such $\r b_{\r i}$ the finite sequence of steps from which  $\r b_{\r i}$ is the final step in deducing that $\rm b_{i}\in C_N(X)$. The resulting finite collections of steps are then renumbered. The final step in this new finite deduction is $\r x.$ Since the reasons for all of the steps is either the original {\bf I} or {\bf RI}, and {\bf RI} contains predetermined n-ary relations that are not dependent upon any deduction, then the finite sequence obtained in this manner is a deduction for a member of $\rm C_N(X)$. Hence,  $\rm x \in C_N(X).$ Consequently,  $\rm C_N(C_N(X)) = C_N(X).$ The finite requirement is obvious since there are only a finite number of steps in any deduction. Note that $\rm C_N(\emptyset) \supset B,$ where $\rm B$ is the set of all $\rm x \in L$ such that $\r x$ is a step obtained only by the rule $\b I.$  Throughout the remainder of this paper,  it is assumed that all ``deductions'' follow these procedures and the corresponding consequence operator is defined as in this paragraph.\parm
\noindent{\bf 4. Intrinsic natural laws or processes.}\parm

For ``scientific deduction'' for a fixed science-community, i,  we need to consider as our rules of inference a collection ${\bf R_i} = {\bf RI}$  of {\bf all} of the ``rules of inference used by this specific scientific-community and allowed by their scientific method'' as they are applied to a  specified language $\Sigma_{\r i},$ the language for ``their science.''  At present, this definition for $\bf R_i$ is rather vague. Hence, the existence of such a set $\bf R_i,$ the rules of inference for a science-community, is an assumption. Of course, as $\Sigma_{\r i}$ changes, so might the $\bf R_i$ be altered. The ${\bf R_i}$ can also change for other valid reasons. From this a specific ``science'' consequence operator $\r S_{\r N_{\r i}}$ is generated for each set of pre-axioms $\r A_{\r i}\cup \r N_{\r i},$ where $\rm A_i$ are the basic logical axioms and $\rm N_i$ the natural laws or processes. For proper application,  the science consequence operator is applied to specific natural-systems, not those generally described. Thus $\r S_{\r N_{\r i}}$ has physical meaning only when $\r S_{\r N_{\r i}}$ is applied to an $\r X$ where every member of $\r X$ and $\r S_{\r N_{\r i}}(\r X)$ is a ``tagged'' statement
that identifies a specific natural-system (Herrmann, 1999). In all that follows,  we are working in a particular $\rm U_{\r i} \subset\Sigma_{\r i}$ of natural laws or processes that are accepted by a particular science-community at this particular moment of time and that are stated using the language $\Sigma_{\r i}.$  \pars 
 
The axiomatic consequence operator $\r S_{\r N_{\r i}}\colon \power {\Sigma_{\r i}} \to \power {\Sigma_{\r i}},$ where $\r S_{\r N_{\r i}}(\emptyset) \supset (\r A_{\r i}\cup \r N_{\r i}),$ can be reduced,  formally,  to an axiomless consequence operator on the language $\Sigma_{\r i} - \r S_{\r N_{\r i}}(\r A_{\r i}\cup \r N_{\r i})$ as shown by Tarski (1930, p. 67). Let $\r V = \{\r A_{\r i},  \r N_{\r i}.\}$ For each $\r X \subset \Sigma_{\r i} - \r S_{\r N_{\r i}}(\r A_{\r i}\cup \r N_{\r i}),$ let $\r S^{\r V}_{\r N_{\r i}}(\r X) =(\Sigma_{\r i}- \r S_{\r N_{\r i}}(\r A_{\r i}\cup \r N_{\r i}))\cap \r S_{\r N_{\r i}}(\r X).$ For this $\r S_{\r N_{\r i}},$ the operator $\r S^{\r V}_{\r N_{\r i}}$ is a consequence operator on $\Sigma_{\r i} - \r S_{\r N_{\r i}}(\r A_{\r i}\cup \r N_{\r i})$ and has the property that $\r S^{\r V}_{\r N_{\r i}}(\emptyset) = \emptyset.$ Thus using $\r S_{\r N_{\r i}}(\r A_{\r i}\cup \r N_{\r i})$ as a set of axioms,  logical and physical,  $\r S^{\r V}_{\r N_{\r i}}$ behaves as if it is axiomless,  where the explicit natural laws or processes $\r N_{\r i}$ behave as if they are implicit. Since,  in general,  $\r S_{\r N_{\r i}}(\r A_{\r i}\cup \r N_{\r i}) \subset \r S_{\r N_{\r i}}(\r X),$ the only consequences that are not but specific deductions from the pre-axioms $\rm A_i \cup N_i$ are members of $\r S_{\r N_{\r i}}(\r X) - \r S_{\r N_{\r i}}(\r A_{\r i}\cup \r N_{\r i}),$ where the explicit $\r X$ should not include the axioms $\r S_{\r N_{\r i}}(\r A_{\r i}\cup \r N_{\r i})$.  
Physically,  $\r S^{\r V}_{\r N_{\r i}}$ is the exact operator that,  using implicitly such axioms as $\rm S_{N_{ i}} (A_{i} \cup N_{i}),$ characterizes the coalescing of a given fundamental collection of named and tagged objects in $\r X$ and that creates a different natural-system or that alters natural-system behavior. The use of axiomless consequence operators is a definite and useful simplification.\pars 
 Applying the above to an entire family of science-communities, we have for an arbitrary science-community, i, a nonempty sequentially represented collection $\rm V_i = \{A_i,  \{N_{ij}\mid j \in \nat\}\}$ such that for any $\rm N_{ij} \in {V_i},$ the set map $\rm S^{V_i}_{N_{ij}}$ defined for each $\rm X \subset (\Sigma_i- (\bigcup \{S_{N_{ij}}(A_i \cup N_j)\mid j \in \nat\})) = \Lambda_i$ by $\rm S^{V_i}_{N_{ij}}(X) =\Lambda _i\cap S_{N_{ij}}(X)$ is a consequence operator defined on $\rm \Lambda_i.$ (The set $\nat$ is the natural numbers not including 0.) The family $\rm V_i$ may or may not be finite. In many cases, it is denumerably since to apply   $\rm S^{V_i}_{N_{ij}}$ to a specifically tagged description $\r X$ certain parameters within the appropriate set of natural laws or processes must be specified so as to correspond to the specific $\r X.$ We assume that the applicable set of natural laws or process $\{\rm N_{ij}\}$ is  the range of a sequence. This will not affect the conclusions since this yields that $\rm V_i$ can be finite or denumerable. Note that for some of the $\rm N_{nm}$ and some tagged $\rm X \subset \Lambda_i$ to which the $\rm N_{nm}$ either does not apply or does not alter, we would have that $\rm S^{V_n}_{N_{nm}}(X) = X.$ For logical consistency, 
it is significant if there exists some type of unifying consequence operator that will unify the separate theories not only applied by a specific science-community (i), but within all of science. \parm
\noindent {\bf 5. An ultralogic unification for all physical theories.}\parm
Although all that follows can be applied to arbitrary science-communities, for notational convenience, consider but one science-community. Thus assume that we have one language for science $\Sigma$ and one sequentially represented countable family of natural laws or processes and logical axioms $\rm A_j \cup  N_j$ as well as one family of sequentially represented rules of inference $\bf R_j$ that generate each specific theory. It is, of course, assumed that ``science,'' in general, is a rational and logically consistent discipline. Let sequentially represented $\rm V = \{A_j \cup N_j\mid \ j \in \nat\}.$ This yields the sequentially represented countable set of all physical theories $\rm \{S_{N_{i}}\mid j \in \nat \}$ and the countable set $\rm \{S^V_{N_{j}}\mid j \in \nat \}$ of intrinsic sequentially represented consequence operators defined on $\rm \Sigma- (\bigcup \{S_{N_{j}}(A_j \cup N_j)\mid j \in \nat\}) = \Lambda$. The following theorem and corollary do not depend upon each member of $\rm \{S_{N_j}\mid j \in \nat\}$ being declared as a ``correct'' physical theory.\pars
Our interest is in the non-trivial application of, at the least, one of these theories to members of $\power {{ \Lambda}}.$\parm
{\bf Definition 5.1.} A nonempty $\rm X \subset {\Lambda}$ is called a {\it significant} member of $\power {\Lambda}$ if there exists some $\rm i \in \nat$ such that $\rm X \not= S^V_{N_i}(X).$ \parm

In what follows, we consider all of the previously defined notions but only with respect to this informal $\r V$ and the language ${\Lambda}.$ Now embed all of these informal results into the formal superstructure ${\cal M} = \langle {\cal N}, \in ,= \rangle$ as done in Herrmann (1987, p. 5; 1993, pp. 9-11). Further, consider the structure $\Hyper {\cal M} = \langle\Hyper {\cal N}, \in,=\rangle$ a nonstandard and elementary extension of $\cal M$ that is a $2^{\vert {\cal M} \vert}$-saturated enlargement ($\vert \cdot \vert$ denotes cardinality). Finally, consider the superstructure ${\cal Y},$ the {\it Grundlegend Structure} (Herrmann, 1993, pp. 22-24). We note that such a structure based upon the natural numbers appears adequate for our analysis since this investigation is only concerned with members of a denumerable language. However, if one wishes to include additional analysis, say with respect to the real numbers, then the {\it Extended Grundlegend Structure} (Herrmann, 1993, p. 70) can be utilized. The approach seems at first to be rather obvious. Simply consider an $\rm W \subset  \Lambda.$ Then the result $\rm \bigcup\{S^V_{N_j}(W)\mid j \in \nat \}$ is most certainly a unification for all of the physical science theories where each theory is represented by a $\rm S_{N_j}.$ However, in general, this union process does not yield a consequence operator (Herrmann, 1987, p. 4). Thus, although each theory may acceptably predict natural-system behavior, such a combined theory may not lend itself to a unification that can be ``rationally'' presented. Consequently, what is sought is a unification that generates each of the results $\rm S_{N_j}(W)$ and this generation is by means of a consequence operator styled process. \parm

{\bf Theorem 5.1.} {\it Given the language $ {\Lambda}$, and the sequentially represented set of consequence operators $\rm \{S^V_{N_j}\mid j \in \nat \}.$ Then there exists an injection ${\cal S}$ on the set $\r M$ of all significant subsets of ${\Lambda}$ into $\Hyper {(\bf {\cal C}_f({\bf \Lambda}))}$ such that for each $\rm W  \in M,$ ${\cal S}_{\rm W}$ is a   nonstandard consequence operator, an ultralogic, such that $\bigcup\{\b S^{\b V}_{\b N_j}({\b W})\mid j \in \nat \}\subset \bigcup\{\Hyper {\b S^{\b V}_{\b N_j}}(\Hyper {\b W})\mid j \in \nat \}= \bigcup\{\Hyper {(\b S^{\b V}_{\b N_j}(\b W))}\mid j \in \nat \}\subset {\cal S}_{\rm W}(\Hyper {\b W})$ and $\bigcup\{{{\b S^{\b V}_{\b N_j}({\b W})}}\mid j \in \nat \}=  {\cal S}_{\rm W} (\Hyper {\b W} ) \cap {\bf \Lambda}.$}\parm 
Proof. In Herrmann (1987, p. 4), a very special set of consequence operators is defined and shown to be closed under the union operator. For this application and for a given $\rm X \in  M$, the set is $\rm H_X = \{C(Y,  X )\mid Y \subset  \Lambda \}.$ Each of the consequence operators in $\rm H_X$ is defined as follows: for each $\rm Z \subset \Lambda,\  C(Y, X )(Z) = Z \cup Y,$ if $\rm Z\cap X \not= \emptyset$; and $\rm C(Y, X )(Z)=Z$ otherwise.  The set $\rm H_X$ is closed under the union operator in the following sense. Consider $\rm \{C(Y_1, X ), \ldots, C(Y_n, X)\},\ n >1; \ Y_k \subset  \Lambda, \ 1 \leq k \leq n.$ Then $\rm C(Y_1 \cup \cdots\cup Y_n, X )(Z) = \bigcup \{C(Y_1, X )(Z), \ldots, C(Y_n, X )(Z)\}= F(Z),\ F \in H_X.$  \pars
Consider the entire set of intrinsic consequence operators $\rm \{S^V_{N_j}\mid j \in \nat \}.$ Define by induction, with respect to the sequentially represented $\rm \{S^V_{N_{i}}\mid j \in \nat \},$
 $\rm C_1(Z) = C(S^V_{N_1}(X),X)(Z),\ C_2(Z)= C(S^V_{N_1}(X) \cup S^V_{N_2}(X), X )(Z),\ldots, C_n(Z)= C(S^V_{N_1}(X) \cup \cdots \cup S^V_{N_n}(X), X )(Z).$ From this definition, it follows that for any $\rm n \in \nat$ the equation (*) $\rm C_n(X)= S^V_{N_1}(X) \cup \cdots \cup S^V_{N_n}(X)$ holds for each $\rm X \subset  \Lambda$. All of the above is now embedded into $\cal M$ and then considered as embedded into the superstructure $\cal Y.$  Since $\rm \{S^V_{N_{i}}\}$ is sequentially represented, there is a fixed sequence $g$ such that $g(i) = {\bf S^V_{N_i}},\  g[\nat] =  \{\b S^{\b V}_{\b N_j}\mid j \in \nat \}$ and $g(i)(\b X) = {\bf S^V_{N_i}}(\b X).$ Hence for arbitrary $\rm X \subset \Lambda$, utilizing  $g$, the above inductive definition yields a sequence $f_{\b X} \colon \nat \to {\bf H_X}$ such that $f_{\b X}(j) = {\bf C_j}$ and $f_{\b X}(j)(\b X) = {\bf C_j}(\b X)$ and, as embedded into $\cal M$, equation (\b *) holds.\pars
Let $\rm X \subset \Lambda.$ Then the following sentence holds in $\cal M$. $$\forall x\forall i((x \in {\bf \Lambda}) \land (i \in \nat) \to  (x \in  f_{\b X}(i)( \b X ) \iff $$ $$ \exists j ((j \in \nat) \land (1\leq j \leq i)\land (x \in g(j)({\bf X})))))\eqno (1)$$\smallskip
\noindent By *-transfer, the sentence
$$\forall x\forall i((x \in 
\hyper{{\bf \Lambda}}) \land (i \in \hypernat) \to  (x \in  \Hyper {(f_{\b X}(i)( \b X ))} \iff $$ $$ \exists j ((j \in \hypernat) \land (1\leq j \leq i)\land (x \in \Hyper {(g(j)({\bf X}))}))))\eqno (2)$$\smallskip
\noindent holds in $\Hyper {\cal M}.$ Due to our method of embedding and identification, sentence (2) can be re-expressed as
$$\forall x\forall i((x \in 
\hyper{{\bf \Lambda}}) \land (i \in \hypernat) \to  (x \in  \hyper {f_{\b X}(i)}( \Hyper {\b X }) \iff $$ $$ \exists j ((j \in \hypernat) \land (1\leq j \leq i)\land (x \in \hyper {g(j)}(\Hyper {{\bf X}})))))\eqno (3)$$\smallskip
\noindent Next consider $\hyper {f_{\b X}} \colon \hypernat \to \Hyper {\bf H_X}$ and any $\lambda \in \hypernat - \nat.$ Then internal $\hyper {f_{\b X}}(\lambda) \in \Hyper {\bf H_X}$ is a nonstandard consequence operator, an ultralogic, that satisfies  statement (3). Hence, arbitrary $j \in \nat$
and $w \in \hyper {g}(j)(\Hyper {\bf X}) = \Hyper {\bf S^V_{N_j}}(\Hyper {\b X})=\Hyper {({\bf S^V_{N_j}(X)})}\subset \hyper {{\bf \Lambda}}$ imply that $w \in \hyper {f_{\b X}(\lambda)}( \Hyper {\b X} )$ since $1 \leq j < \lambda.$ Observe that ${}^{\sigma} {({\bf S^V_{N_j}(X)})}\subset \Hyper {({\bf S^V_{N_j}(X)})}.$ However, under our special method for embedding ${}^{\sigma} ({\bf S^V_{N_j}(X)}) = {\bf S^V_{N_j}(X)},$ for an arbitrary $\bf X \subset {\bf \Lambda}.$\pars
The final step is to vary the $\rm X \in M.$ We first show that for two distinct $\rm X,\ Y \in {\r M}$ there is an $\rm m \in \nat$ such that 
$\rm C_m^X = C(S^V_{N_1}(X) \cup \cdots \cup S^V_{N_m}(X), X) \not= C_m^Y = C(S^V_{N_1}(Y) \cup \cdots \cup S^V_{N_m}(Y), Y).$ Since $\rm X,\ Y$ are nonempty, distinct and arbitrary, we need only assume that there is some $\rm x \in X - Y.$ Hence there is some $\rm i \in \nat$ and $\rm j \in \nat$ such that $\rm X \subset S^V_{N_i}(X) \not= X$ and  $\rm Y \subset S^V_{N_j}(Y) \not= Y.$
Consider some $\rm m \in \nat$ such that $\rm i,\ j \leq m.$ Then 
$\rm C_m^X(\{x\}) = C(S^V_{N_1}(X) \cup \cdots \cup S^V_{N_m}(X), X)(\{x\})= 
S^V_{N_1}(X) \cup \cdots \cup S^V_{N_m}(X) \not= X\subset C_m^X(\{x\}).$ But $\rm C_m^Y(\{x\}) = C(S^V_{N_1}(Y) \cup \cdots \cup S^V_{N_m}(Y), Y)(\{x\}) = \{x\} \not=  C_m^X(\{x\})$ Thus $\rm C_m^Y(\{x\})\not=  C_m^X(\{x\}).$ Further, for any ($\dagger$) $\rm k \in \nat,\ m \leq k,\  C_k^Y(\{x\})\not=  C_k^X(\{x\}).$ Consider these results formally stated. Then  by *-transfer, for each distinct pair $\b X,\ \b Y \in \b M$ there exists some $m \in \hypernat$ such that $\hyper {f_{\b X}}(m) \not= \hyper {f_{\b Y}}(m).$ Thus for $\b X, \ \b Y \in \b M, \  \b X \not= \b Y,$  $A(\b X, \b Y) =\{m 
\mid (m \in \hypernat)\land \hyper {f_{\b X}}(m) \not= \hyper {f_{\b Y}}(m)\}$ is nonempty.  
We use the Axiom of Choice for the  general set theory (Herrmann, 1993, p. 2) used to construct our $\cal Y.$ Hence, there exists a set $B,$ within our structure, containing one and only member from each $A(\b X, \b Y).$\pars
The internal binary relation $\{(x,y)\mid (x \in \hypernat)\land (y \in \hypernat)\land  (x \leq y)\}$ is from *-transfer of $\nat$ properties a concurrent relation with respect to the range $\hypernat$. Since $\Hyper {\cal M}$ is  a $2^{\vert {\cal M} \vert}$-saturated enlargement and $\vert B \vert < 2^{\vert {\cal M} \vert}$, there is some $\lambda \in \hypernat$ such that for each $i \in B,\ i \leq \lambda$. Considering this $\lambda$ as fixed, then by *-transfer of $(\dagger)$, it follows that for any distinct $\rm X ,\ Y \in M$ $\hyper {f_{\b X}}(\lambda) \not= \hyper {f_{\b Y}}(\lambda).$ Since $\r M$ is injectively mapped onto $\b M$, there exists an injection $\cal S$ on the set $\r M$ such that each $\rm W \in M$, $ {\cal S}_{\rm W}= \hyper {f_{\b W}}(\lambda) \in  \Hyper {({\bf {\cal C}_f(\Lambda)})}.$   Considering the general properties for such an $\hyper {f_{\b W}}(\lambda)$ as discussed above, it follows, that $\bigcup\{{{\b S^{\b V}_{\b N_j}({\b W})}}\mid j \in \nat \}\subset {\cal S}_{\rm W} (\Hyper {\b W} ) \cap {\bf \Lambda}.$\pars
Now assume that standard $\b a \in {\cal S}_{\rm W} (\Hyper {\b W} )- \bigcup\{{{\b S^{\b V}_{\b N_j}({\b W})}}\mid j \in \nat \}.$ (For our identification and embedding, $\hyper {\b a} = \b a.$) Then the following sentence
$$ \forall x \forall i( (x \in  {\bf \Lambda}) \land (i \in \nat) \land x \in g(i)(\b W) \to x \not= \b a) \eqno (4)$$
holds in $\cal M$ and, hence,
$$ \forall x \forall i( (x \in \hyper{ {\bf \Lambda}}) \land (i \in \hypernat) \land x \in \hyper {g(i)}(\Hyper {\b W}) \to x \not= {\b a}) \eqno (5)$$
holds in $\Hyper {\cal M}.$ But since $\b a \in \hyper {f_{\b W}}(\lambda)(\Hyper {\b W}),$ then statement (5) contradicts statement (3) and the proof is complete.\qed
{\bf Corollary 5.1.1} {\it If $\rm \{S^V_{N_j} \mid j \in \nat\}$ represents all of the physical theories that describe natural world behavior, then the choice function and the last equation in Theorem 5.1 correspond to an ultralogic unification for $\rm \{S^V_{N_j} \mid j \in \nat\}$}.\parm  

Note that usually $\r W$ is a finite set. Assuming this case, then again due to our method of embedding $\Hyper {{\b W}} = \b W.$ In statement (3), $\Hyper {g(i)} = \Hyper {{\bf  S^V_{N_i}}}.$ However, $\rm S^V_{N_i}$ has had removed all of the steps that usually yield an infinite collection of results when $\rm S_{N_i}$ is applied to $\r W.$ Thus, in most cases, 
$\rm S_{N_i}(W)$ is a finite set. Hence, if one assumes these two finite cases, then we further have that ${\bf S^V_{N_j}(W)}= \Hyper {({\bf S^V_{N_j}(W)})}.$ However, each ${\cal S}_{\rm W}$ remains a nonstandard ultralogic since each ${\cal S}_{\rm W}$ is defined on the family of all internal subsets of $\hyper { {\bf \Lambda}}$ since the consistency of the combined collection of all of the scientific theories implies that $\Lambda$ is denumerable. Of significance is that corollary 5.1.1 is technically falsifiable. The most likely falsifying entity would be the acceptance of a physical theory that does not use the rules of inference as setout in section 3. In particular, when different hypotheses are considered, the requirement that the rules of inference {\bf RI} cannot be altered.  \pars
 Such operators as ${\cal S}_{\rm W}$ can be interpreted in distinct ways. If they are interpreted in a physical-like sense, then they operator in a region called the {\it nonstandard physical world} (Herrmann, 1989), where $\rm W$ corresponds physically to the natural-system it describes. The restriction ${\cal S}_{\rm W}( {\Hyper {\b W}} ) \cap { {\bf \Lambda}}$ then represents a natural world entity. As a second interpretation, $\cal S$ would represent an intrinsic process that appears to guide the development of our universe and tends to verify the Louis de Broglie statement. ``[T]he structure of the material universe has something in common with the laws that govern the workings of the human mind'' (March, 1963, p. 143). 
 \parm
\noindent {\bf 6. Probability models.}\parm
In Herrmann (1999, 2001), it is shown that given a specific probability theory for a specific source or natural-system described by a single sentence $\{\r G\}$ that predicts that an event $\r E$ will occur with probability $p$ then there is an ultralogic $P_p$ that generates an exact sequence of such events the relative frequency of which will converge to $p.$ It is also shown that the patterns produced by the frequency functions for statistical distributions that model natural-system behavior are also the results of applications of ultralogics. Although the main results in these papers state as part of the hypothesis that $p$ is theory predicted, the results also hold if $p$ or the distribution is obtained from but empirical evidence. Theorem 2 in Herrmann (1999, 2001) actually corresponds to Theorem 5.1. Notice that throughout Theorem 2 in Herrmann (1999,2001), the singleton set $\{\r G\}$ can be replaced by any nonempty $\rm W \subset {\Lambda},$ where the $H$ is defined as in this paper, and not only does this Theorem 2 still hold but so do the results on distributions. \pars
Are these results for probability models consistent with Theorem 5.1?
If probability models predict natural-system behavior, in any manner, then, in general, the natural laws or processes $\r N$ that are assumed to lead to such behavior only include a statement that claims that the event sequences or distributions appear in the natural world to be ``randomly'' generated. It is precisely the results in Herrmann (1999, 2001) that show that in the nonstandard physical world such behavior need not be randomly obtained but can be specifically generated by ultralogics. These results are thus consistent since the ultralogics obtained from Theorem 2 neither correspond to nor apply to any nonstandard extension of the notion of standard ``randomness.''\parm     

\centerline{\bf References}\parm

\id{F}erris,  Timothy. (1977), {\it The Red Limit.}  New York: Bantam Books.\smallskip
\id{H}errmann, Robert A. (2001), ``Ultralogics and probability models,'' {\it International Journal of Mathematics and Mathematical Sciences} (To appear).\smallskip
\id{H}errmann, Robert A. (1999), ``The wondrous design and non-random character of `chance' events,'' http://www.arXiv.org/abs/physics/9903038\smallskip
\id{H}errmann, Robert A. (1993),  {\it The Theory of Ultralogics.} \hfil\break http://www.arXiv.org/abs/math.GM/9903081 and/9903082\smallskip

\id{H}errmann, Robert A. (1989), ``Fractals and ultrasmooth microeffects.'' {\it Journal of Mathematical Physics} 30(4):805-808. \smallskip

\id{H}errmann, Robert A. (1987), ``Nonstandard consequence operators''. {\it Kobe Journal of  Mathematics} 4:1-14. http://www.arXiv.org/abs/math.LO/9911204\smallskip

\id{M}arch, Arthur and Ira M. Freeman. (1963), {\it The New World of Physics}. New York: Vintage Books.\smallskip

\id{T}arski,  Alfred. (1956), {\it Logic, Semantics, Metamathematics; papers from 1923 - 1938},  Translated by J. H. Woodger$.$ Oxford: Clarendon Press.

\end